\documentclass[a4paper]{article}

\usepackage{INTERSPEECH2022}
\usepackage{makecell}
\usepackage{threeparttable}
\usepackage{multirow}
\usepackage{verbatim}
\usepackage{enumitem}
\usepackage{amsfonts} 
\usepackage{subfig}
\usepackage{hyperref}

\usepackage{breakurl}
\makeatletter
\def\UrlAlphabet{%
      \do\a\do\b\do\c\do\d\do\e\do\f\do\g\do\h\do\i\do\j%
      \do\k\do\l\do\m\do\n\do\o\do\p\do\q\do\r\do\s\do\t%
      \do\u\do\v\do\w\do\x\do\y\do\z\do\A\do\B\do\C\do\D%
      \do\E\do\F\do\G\do\H\do\I\do\J\do\K\do\L\do\M\do\N%
      \do\O\do\P\do\Q\do\R\do\S\do\T\do\U\do\V\do\W\do\X%
      \do\Y\do\Z}
\def\UrlDigits{\do\1\do\2\do\3\do\4\do\5\do\6\do\7\do\8\do\9\do\0}
\g@addto@macro{\UrlBreaks}{\UrlOrds}
\g@addto@macro{\UrlBreaks}{\UrlAlphabet}
\g@addto@macro{\UrlBreaks}{\UrlDigits}
\makeatletter
\title{ATST: Audio Representation Learning with Teacher-Student Transformer}
\name{Xian Li, Xiaofei Li$^{*}$\thanks{* corresponding author}}
\address{$^{1}$Westlake Institute for Advanced Study \& $^{2} $Westlake University  , Hangzhou, China}
\email{lixian@westlake.edu.cn, lixiaofei@westlake.edu.cn}

\begin{document}

\maketitle
\begin{abstract}
Self-supervised learning (SSL) learns knowledge from a large amount of unlabeled data, and then transfers the knowledge to a specific problem with a limited number of labeled data. SSL has achieved promising results in various domains. This work addresses the problem of segment-level general audio SSL, and proposes a new transformer-based teacher-student SSL model, named ATST. A transformer encoder is developed on a recently emerged teacher-student baseline scheme, which largely improves the modeling capability of pre-training.
In addition, a new strategy for positive pair creation is designed to fully leverage the capability of transformer. Extensive experiments have been conducted, and the proposed model achieves the new state-of-the-art results on almost all of the downstream tasks.  

\end{abstract}

\section{Introduction}
\label{sec:intro}
%With the rapid development of deep learning, great progress has been achieved in various domains.  However, it usually needs a large mount of labeled data to train a well-performed model for each specific problem.
%In fact, data labeling is usually with high cost. To address this problem, self-supervised learning(SSL) emerges, which learns knowledge from large amount of unlabeled data, and then transfers the knowledge to a specific problem with a limitefd number of labeled data.  

Recently, learning audio representations with self-supervised learning (SSL) has been widely studied  \cite{oord_representation_2019}\cite{tagliasacchi_pre-training_2020}\cite{saeed_contrastive_2020}\cite{fonseca_unsupervised_2021}\cite{niizumi_byol_2021}. Among them, the contrastive learning methods \cite{saeed_contrastive_2020}\cite{fonseca_unsupervised_2021}\cite{niizumi_byol_2021} maximize the classification similarity of two augmented views of the same audio clip (called a positive pair), having shown a great promise for learning good representation. The above idea often confronts the issue of model collapse, e.g. the model can find an easy solution to output a constant value for any inputs. COLA \cite{saeed_contrastive_2020} and  \cite{fonseca_unsupervised_2021} overcome this issue by distinguishing positive audio samples from a batch of negative audio samples. Considering the fact that for audio data, negative samples are possibly similar to positive samples in some scenarios, BYOL-A \cite{niizumi_byol_2021} proposed to discard the negative samples, and use a teacher-student scheme to overcome the issue of model collapse. The teacher and student networks process two different views of an audio clip, and the student network is trained to predict a representation being identical to the prediction of the teacher model. The teacher network is updated by taking an exponential moving average (EMA) of the student network. Another technical line for audio SSL follows the spirit of Bert \cite{devlin_bert_2019}, performing a predictive task for the masked frames, e.g. wav2vec2 \cite{baevski_wav2vec_2020} and Hubert \cite{hsu_hubert_2021}. 

Transformer network has shown powerful abilities in learning long-term dependencies, and has been used for speech SSL in several works, e.g. MockingJay \cite{liu_mockingjay_2020}, wav2vec2 \cite{baevski_wav2vec_2020}, Hubert \cite{hsu_hubert_2021}, Tera \cite{liu_tera_2021}. As for general audio SSL, people usually use convolution neural network (CNN), e.g. in COLA, BYOL-A, etc. To the best of our knowledge, SSAST \cite{gong_ssast_2022} and \cite{srivastava_conformer-based_2022} are the only two  very recent works that use transformer for general audio SSL. They both follow the line of wav2vec2 \cite{baevski_wav2vec_2020}. 

According to the grain size of the representation at the pre-training stage, all the above methods can be categorized into two types: segment-level method and frame-level method. The segment-level methods, e.g. COLA \cite{saeed_contrastive_2020} and BYOL-A \cite{niizumi_byol_2021}, extract a fixed-length segment embedding from an input audio segment. On the other hand, the frame-level methods, e.g. SSAST \cite{gong_ssast_2022} and \cite{srivastava_conformer-based_2022}, extract an individual embedding for all frames. Learning a segment embedding is suitable for a variety of segment-level audio tasks, e.g. sound event classification, music instrument classification, speaker identification, etc. COLA \cite{saeed_contrastive_2020} and BYOL-A \cite{niizumi_byol_2021} have been proven very effective for segment-level general audio SSL, and have achieved the state-of-the-art performance. Although SSAST \cite{gong_ssast_2022} and \cite{srivastava_conformer-based_2022} are pre-trained by a frame-level criterion, the downstream tasks that they have applied to are all segment-level tasks, where average pooling is applied to obtain the segment embedding.  
%Also, SIMCLR\cite{chen_simple_2020} and BYOL\cite{grill_bootstrap_2020}, where COLA and BYOL-A get inspiration from respectively, have proved effectiveness of learning at global-image level for image classification problems.   % In intuitions, */

This work focuses on the problem of segment-level general audio SSL, and proposes a new transformer-based teacher-student SSL model, named ATST \footnote{code: \url{https://github.com/Audio-WestlakeU/audiossl}}. Main contributions of this work include: i) adopting transformer encoder into the baseline teacher-student scheme of BYOL-A \cite{niizumi_byol_2021}, which shows a clear superiority over the CNN encoder of BYOL-A, especially for learning the long-term semantic information of speech; ii) proposing a new view creation strategy. BYOL-A uses one short segment to create two views (one positive pair). Instead, we propose to use two different long segments, which is more fit for transformer, as the network needs to learn longer temporal dependencies and to match a more distinct positive pair created by two segments. The length of segments is carefully studied to control the distinction and overlap of the two segments, which is especially important for rationalizing the difficulty of matching positive pairs. Experiments have been conducted using the large-scale Audioset \cite{gemmeke_audio_2017} dataset for pre-training. Downstream tasks cover all the three types of audio signals, namely audio event, speech, and music. Ablation experiments show the effectiveness of each of the proposed modules. The proposed model as a whole achieves the new state-of-the-art results on almost all of the downstream tasks, and surpasses other methods by a large margin on some of the downstream tasks. For example, the accuracy of speaker identification is 72\% versus 40.1\% without finetuning, and  94.3\% versus 80.8\% after finetuning.

\section{The Proposed Method}
\label{sec:method}

%We exploit the Transformer architecture and BYOL to learn segmental-level representation for audio signal, as is depicted in Figure \ref{fig:framework}. 

\subsection{Baseline Teacher-Student Scheme  }
\label{sec:baseline}

 In this work, we adopt the teacher-student scheme as our baseline framework, which was first proposed by Bootstrap you own latent (BYOL) \cite{grill_bootstrap_2020} for image pre-training, and adopted by BYOL-A \cite{niizumi_byol_2021} for audio pre-training. Given one augmented view of an audio clip, the student network is trained to predict a data representation being identical to the teacher network's prediction on one another augmented view of the same audio clip. During training, the teacher network is updated by taking the EMA of student network. Specifically, the student network, defined by a set of weights $\theta$, contains an encoder $f_{\theta}$, a projector $g_{\theta}$ and a predictor $q_{\theta}$, while the teacher network, defined by a set of weights $\phi$, contains only an encoder $f_{\phi}$ and a projector $g_{\phi}$.  The encoders extract a representation from the augmented views. 
 %The projectors are used to improve the representation quality \cite{chen_simple_2020}. 
 It has been shown that the additional predictor in the student network combined with the stop-gradient operation introduced by using EMA teacher network is the key factor that prevents the model from collapsing \cite{chen_exploring_2020}. During training, $\phi$ is updated by the EMA of $\theta$ as: $\phi \leftarrow m\phi + (1-m)\theta$, where $m$ is a decay rate. $\theta$ is updated as follows. Let $(\Vec{X},\Vec{X}^{'} )$ be a pair of positive views created from an audio clip.  $\Vec{X}$ is fed into the teacher network to obtain $\Vec{h}=f_{\phi}(\Vec{X})$ and $\Vec{z}=g_{\phi}(\Vec{h})$. $\Vec{X^{'}}$ is fed into the student network to obtain $\Vec{h^{'}}=f_{\theta}(\Vec{X}^{'})$, $\Vec{z}^{'}=g_{\theta}(\Vec{h}^{'})$ and $q_{\theta}(\Vec{z}^{'})$. $\Vec{z}$ and $q_{\theta}(\Vec{z}^{'})$ are then L2-norm normalized, and the mean square error between them is calculated as $ L_{\theta}$.
 %as $\overline{\Vec{z}}$ and $\overline{q}_{\theta}( \Vec{z}^{'} )$, respectively.
 %$\overline{\Vec{z}} = \Vec{z}/\Vert \Vec{z} \Vert_2$ and $\overline{q}_{\theta}( \Vec{z}^{'} )=q_{\theta}(\Vec{z}^{'})/\Vert  q_{\theta}(\Vec{z}^{'}) \Vert_2$, , where $\Vert \Vert_2$ denotes L2-norm. 
 %The mean square error between them is calculated as $ L_{\theta}$. 
%\begin{equation}
%    L_{\theta}= \Vert \overline{\Vec{z}}-\overline{q}_{\theta}(\Vec{z}^{'}) \Vert_2^2
%\end{equation}
 A symmetric loss $L_{\theta}^{'}$ is also calculated by feeding $\Vec{X}$ to the student network and $\Vec{X}^{'}$ to the teacher network. During training, $\theta$ is updated by minimizing $L_{\theta}^{total} = L_{\theta} + L_{\theta}^{'} $. 
 
In BYOL-A, encoder is a CNN, projectors and predictors are multi-layer perceptrons (MLPs) that consist in a linear layer (with output dimension of 4096) followed by batch normalization, rectified linear units (RELU), and a final linear layer (with output dimension of 256).

%As mentioned in section \ref{sec:intro}, the contrastive method is prone to model collapse. Although BYOL has addressed this issue,  we find appropriate values of learning rate and  $m_0$  are required for the model to learn a uniform representation. To address this, we monitor standard deviation of $z$ during training to find 

\subsection{The Proposed Method}
\begin{figure}[t]
  \centering
  \subfloat[]{\includegraphics[width=\linewidth]{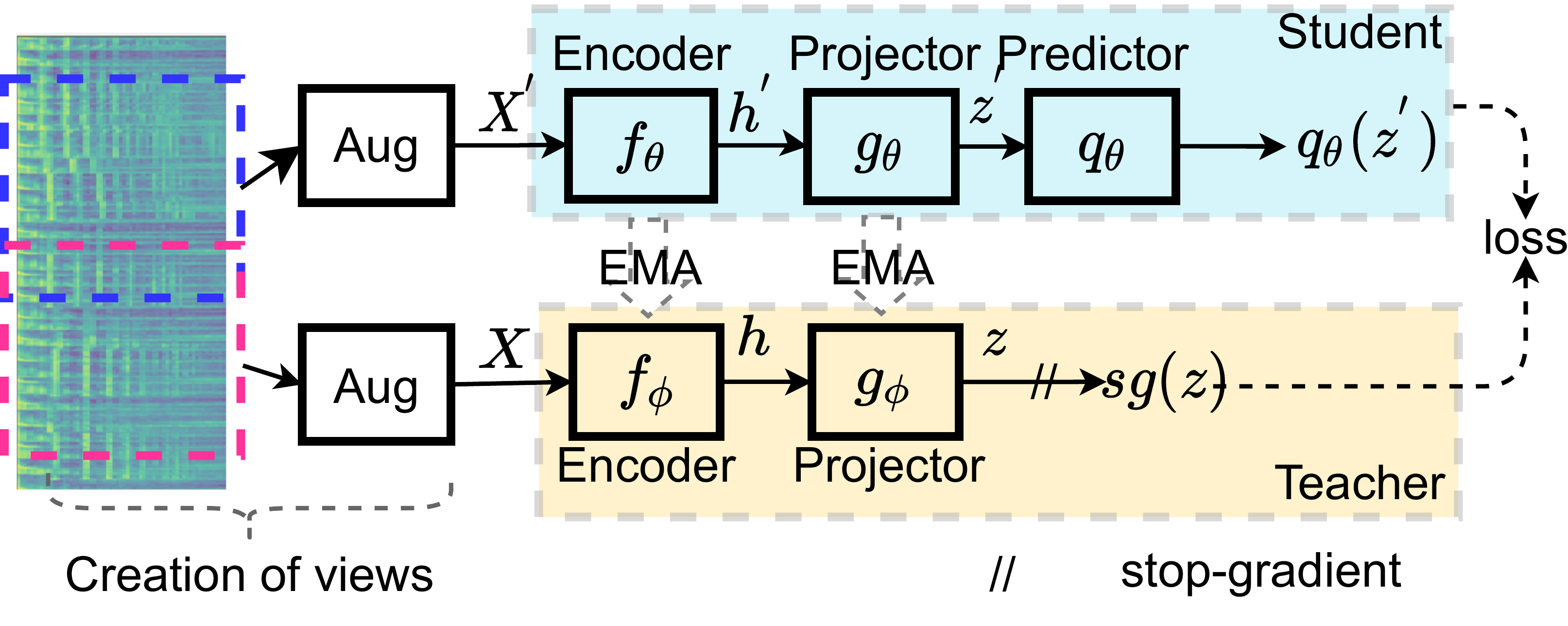}}
 
 \subfloat[]{\includegraphics[width=0.8\linewidth]{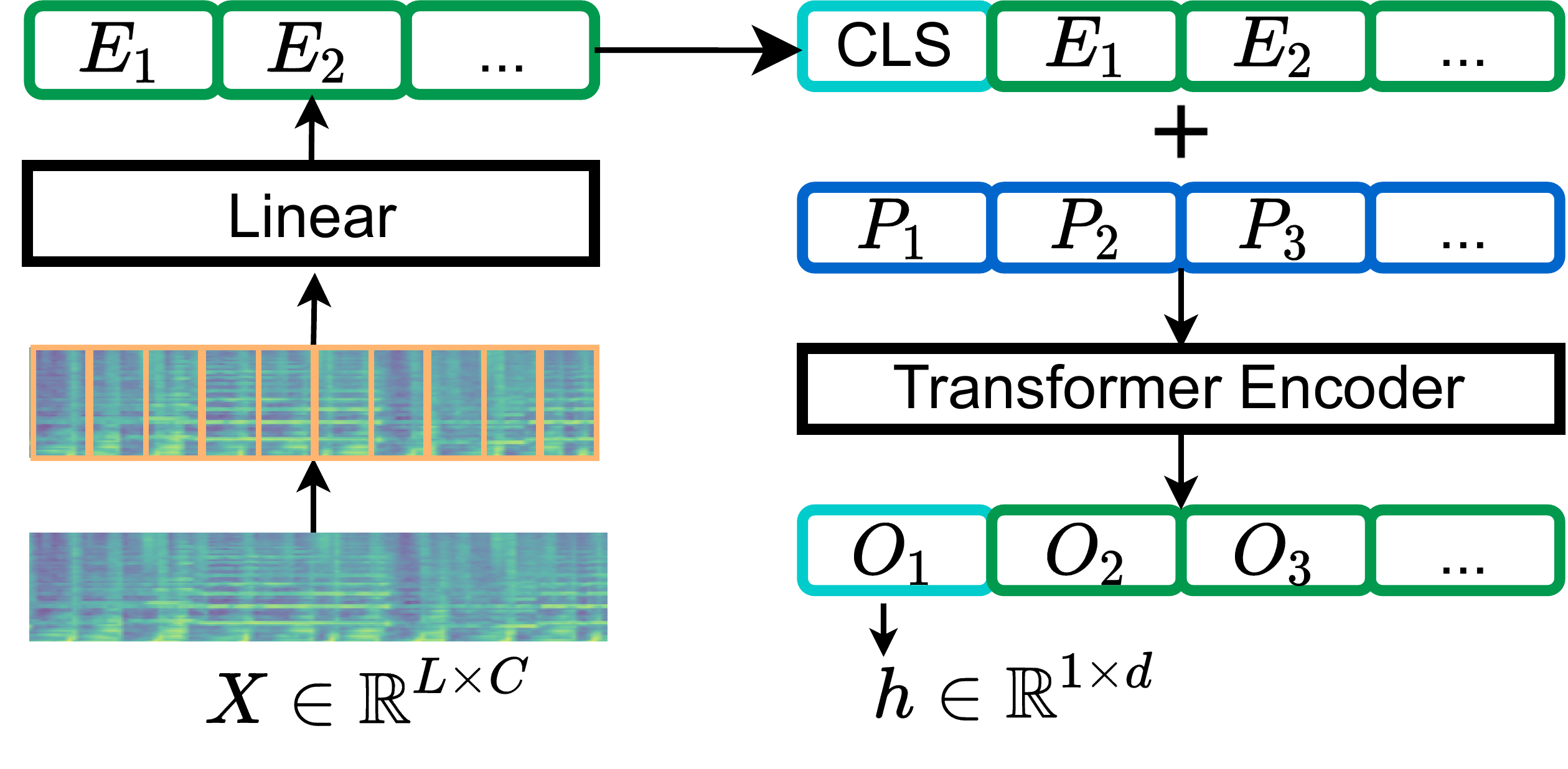}}

  \caption{(a) Overview of the proposed method. "Aug" denotes augmentation (b) Transformer encoder.}
  \label{fig:framework}
 \vspace{-0.6cm}
\end{figure}

Overview of the proposed method is depicted in Fig.~\ref{fig:framework} (a).
The major differences between the proposed method and BYOL-A are two folds. The proposed method uses a transformer as encoder to leverage its powerful abilities on modeling long-term dependencies, and uses a new view (positive pair) creation strategy specifically being fit for the transformer encoder. 

%The data for pretraining is large-scale Audioset\cite{gemmeke_audio_2017} dataset. The full Audioset(\textbf{AS-2M}) contains 200 millions audio clips with length of 10 seconds from Youtube Videos. Besides, we create a subset  with 200 thousands audio clips(\textbf{AS-200K}) by randomly sampling from \textbf{AS-2M}. Note the label of Audioset is not used at pretraining stage. 

\vspace{-0.2cm}
\subsubsection{Creation of Views}
BYOL-A \cite{niizumi_byol_2021} randomly crops a single 1-second segment from the input audio and then creates two views by applying different data augments to this single segment. It is considered in BYOL-A \cite{niizumi_byol_2021} that different segments may be too different to be identified as a positive pair. \cite{fonseca_unsupervised_2021} uses two segments to create positive views, however, it uses negative views to mitigate the problem caused by using two segments.   

Our view creation strategy is shown in Fig.~\ref{fig:framework} (a). 
The time domain input audio clip is first transformed to mel-spectrogram. We randomly crop two different segments from the mel-spectrogram. Then, two types of data augmentation are applied to each of the segments, including Mixup \cite{niizumi_byol_2021} and Random Resized Crop (RRC) \cite{niizumi_byol_2021}, creating two views of the input audio clip, i.e. $(\Vec{X},\Vec{X}^{'})$. Note that this work does not use negative samples. 
In order to take full advantage of the transformer's ability in modeling long-term dependencies, the proposed method intends to use longer segments, e.g. 6 seconds in experiments. The proposed method separately creates two views from two different segments for the purpose of increasing the difficulty of identifying the two views as a positive pair, thus leading the model to learn more generalized representations. On the other hand, the two segments cannot be too far away from each other, otherwise the similarity between them is completely lost. This is guaranteed by properly setting the segment length to make the two segments have a certain portion of overlap. 
Overall, the proposed strategy does not lose the rationality of identifying two segments as a positive pair due to the overlap constraint, and meanwhile increases the task difficulty and thus the model capability by using two segments. 

 %Note that in BYOL-A, due to the temporal cropping performed by RRC, two views created from the single segment can also be seen as two shorter segments with different time spans. However, the two views created by BYOL-A are more temporally closer to each other than ours and share the same semantic content at most of the time. 
 %Note that in computer vision, two crops with different sizes \cite{chen_simple_2020}\cite{grill_bootstrap_2020}\cite{caron_emerging_2021} are usually used, thus encouraging to learn a local to global relationship. However, in our preliminary experiments, the proposed method achieves better result by using two segments with the same length. 
%
\subsubsection{Transformer Encoder}
\label{sec:encoder}

The encoding procedure is illustrated in Fig.~\ref{fig:framework} (b).
The augmented mel-spectrogram $\Vec{X} \in \mathbb{R}^{L\times C}$, where $L$ and $C$ denote frames and channels respectively, is fed into a transformer encoder \cite{vaswani_attention_2017} to obtain a fixed-length segment embedding $\Vec{h} \in \mathbb{R}^{1 \times d} $, where $d$ denotes the dimension of embedding. 

Four consecutive frames of $\Vec{X}$ are first stacked to reduce the temporal resolution and sequence length. The stacked frames are fed to a linear projection layer (with output dimension of $d$) to obtain a new embedding sequence $\Vec{E}\in \mathbb{R}^{\frac{L}{4}\times d}$ as the input sequence of transformer encoder. Besides this input sequence, we use an extra trainable class token $\Vec{CLS} \in \mathbb{R}^{1\times d}$ to represent the entire segment, which is inserted to the beginning of the input sequence. This kind of segment class token is widely used for sentence embedding in neural language processing \cite{devlin_bert_2019}, global image embedding \cite{caron_emerging_2021}, as well as audio segment embedding  \cite{gong_ast_2021}. A trainable absolute  lookup table positional embedding $\Vec{P} \in \mathbb{R}^{(\frac{L}{4}+1) \times d} $  is then added to the input sequence. Eventually, we use a standard transformer encoder \cite{vaswani_attention_2017} to process the input embedding sequence, obtaining an output embedding sequence of $\Vec{O} \in \mathbb{R}^{(\frac{L}{4}+1)\times d}$. In the output sequence, the segment class token, i.e. $\Vec{O_1}$, aggregates information from the embedding sequence at each block of the transformer, based on the self-attention mechanism. Therefore, $\Vec{O_1}$ is taken as the final segment embedding, and is denoted as $\Vec{h}$ or $\Vec{h}^{'}$ in the teacher-student pre-training scheme. 

As for downstream tasks, the pre-trained transformer encoder of teacher network is used as the feature extractor, while the projector is removed.

\section{Experiments}

\label{sec:result}
\begin{table*}[htb]
  \caption{Linear evaluation results of \textbf{Small} model w.r.t. different view creation strategies. "Average" is taken over the last four tasks.}
    \label{tab:segments}
   \centering
  \begin{tabular}{l|cc|ccccc|c}
    \toprule
   \textbf{Method} &\textbf{Segments} & \textbf{\makecell{length of\\segment (s)}}   &   \textbf{\makecell{AS-20K\\mAP}}    & \textbf{\makecell{SPCV2\\Acc (\%)}}    & \textbf{\makecell{VOX1\\Acc (\%)}}   
     & \textbf{\makecell{NSYNTH\\Acc (\%)}} & \textbf{\makecell{US8K\\Acc (\%)}} &\textbf{\makecell{Average\\Acc (\%)}}  \\ 
    \midrule
    \textbf{BYOL-A \cite{niizumi_byol_2021}} & single & 1 & - & 92.2 & 40.1& 74.1& 79.1&  71.4 \\
    \midrule
    \textbf{\multirow{4}{*}{Small (Ours)}}   &     single & 1 & 0.210  & 94.3 &  52.3  & 73.8 & 79.3  & 74.9     \\ 
    &two  & 1 &  0.191  & 91.3 &  50.0 & 74.3 & 76.6 &   73.1   \\ 
    &single  & 6 & 0.257 & \textbf{94.0} &  $57.3 $  &  $73.8$    & 80.9  & 76.5                              \\
     & two & 6 & \textbf{0.279}  & 93.6 &  \textbf{61.9}  & \textbf{75.3} &  \textbf{82.0} &  \textbf{78.2}   \\   
    \bottomrule 

  \end{tabular}
  
\end{table*}

We evaluate the performance of our model under the protocol of linear evaluation or finetuning.  In linear evaluation, the pre-trained encoder is frozen as a feature extractor, on top of which a linear classifier is trained. Whereas in finetuning, the pre-trained encoder and linear classifier are finetuned together.  

\subsection{Implementation Details of Pre-training}

We use Audioset \cite{gemmeke_audio_2017} for pre-training. The full Audioset (\textbf{AS-2M}) contains 200 million audio clips with a length of 10 seconds captured from Youtube Videos. Using the full Audioset, a \textbf{Base} model is trained, which contains 12 blocks, and 12 heads for each block. The dimension and inner dimension are 768 and 3072 respectively. Besides, using a subset with 200 thousand audio clips (\textbf{AS-200K}) randomly sampled from \textbf{AS-2M}, we also trained a \textbf{Small} model, which contains 12 blocks, and 6 heads for each block. The dimension and inner dimension are 384 and 1536 respectively.

\label{sec:impdetail_pre}
 Audio is re-sampled to 16 kHz. Audio clips are transformed to the mel-spectrogram domain, with a Hamming window, a window length of 25 ms, a hop size of 10 ms, and 64 frequency bins ranging from 60 Hz to 7800 Hz. The mel-spectrogram feature is min-max normalized, where the minimum and maximum values are calculated globally on the pre-training dataset. We intentionally set the length of two segments (for creating two views) to 6 seconds, which will leads to a segment overlap of at least 1 second, considering that the length of audio clip is 10 seconds. The two randomly sampled segments are augmented by Mixup and RRC with the same configurations used in BYOL-A \cite{niizumi_byol_2021}.

We pre-train our models with the ADAMW optimizer \cite{loshchilov2017fixing}. The learning rate $lr$ is warmed up for 10 epochs, and then annealed to 1e-6 at cosine rate. Following DINO \cite{caron_emerging_2021}, the weight decay of transformer is increased from 0.04 to 0.4 at cosine rate. The EMA decay rate $m$ increases from an initial value $m_0$ to 1 at cosine rate.  Batch size is set to 1536. The \textbf{Base} model is trained using \textbf{AS-2M} for 200 epochs, with $lr$ being 2e-4, and $m_0$ being 0.9995.
The \textbf{Small} model is trained using \textbf{AS-200K} for 300 epochs, with $lr$ being 5e-4, and $m_0$ being 0.99.
%We set a smaller $lr$ and a larger $m_0$ for the \textbf{Base} model, since we found that adjusting the value of $lr$  and $m_0$ according to the number of training steps is essential to achieve a good performance. Smaller $lr$ and larger $m_0$ are required when training with more steps, otherwise the performance will drop. 

\subsection{Downstream Tasks}
\label{sec:dataset}

Evaluations are carried out on a variety of downstream tasks, which cover all the three types of audio signals, namely audio event, speech and music. Datasets and downstream tasks are described as follows.
\begin{itemize}[leftmargin=*,itemsep=0pt]
\item \textbf{AS-20K} for multi-label sound event classification. We use the balanced subset of Audioset-2M, with 527 audio classes. It contains 20,886 audio clips for training. For test, we use the evaluation set of Audioset, with 18,886 audio clips. 
\item \textbf{US8K} for single-label audio scene classification. We use the Urbansound8k dataset \cite{salamon2014dataset} to classify audio clips (less than 4 seconds) into 10 classes. It contains 8,732 audio clips and has ten folds for cross-validation.
\item \textbf{SPCV2} for spoken command recognition. We use Speech Command V2 \cite{warden2018speech} to recognize 35 spoken commands for one second of audio. It contains 84,843, 9,981 and 11,005 audio clips for training, validation and evaluation, respectively.
\item \textbf{VOX1} for speaker identification. We use the Voxceleb1 dataset \cite{nagrani2017voxceleb}, with 1,251 speakers. It contains 13,8361, 6,904 and 8,251 for training, validation and evaluation, respectively.
\item \textbf{NSYNTH} for music instrument classification. We use the NSYNTH dataset \cite{engel2017neural}, to recognize 11 instrument family classes from 4-seconds audio clips.
\end{itemize}

%\subsubsection{Implemetation Details of Downstream Tasks}
%\label{sec:impdetail_down}
The mel-spectrogram feature is computed in the same way as for the pre-training data.  For linear evaluation, from the pre-trained encoder, segment embedding is obtained by concatenating the class token $\Vec{O_1}$ and the average of the embedding sequence of all blocks. For finetuning, segment embedding is obtained by concatenating the class token and the average of the embedding sequence of the last block. Audio clips that are longer than 12 seconds are centrally cropped with a maximum length of 12 seconds, and then split into 6-second long chunks without overlap. The chunks are independently processed by the pre-trained encoder, and their outputs are averaged to obtain the final segment embedding. Audio clips that are shorter than 6 seconds are directly processed by the pre-trained encoder to obtain the segment embedding.  

 For linear evaluation, we train the linear classifier for 100 epochs with the SGD optimizer. The learning rate is annealed to 1e-6 at cosine rate during training. The optimal initial learning rate is searched for each task separately. Batch size is set to 1024. Augmentation is not used. 
 
 For finetuning, we finetune all models with the SGD optimizer.  The learning rate $lr$ is warmed up for 5 epochs, and then annealed to 1e-6 at cosine rate. The optimal $lr$ is searched for each task separately. Batch size is set to 512. We trained \textbf{SPCV2} and \textbf{VOX1} for 50 epochs, and \textbf{AS-20K} for 200 epochs. For \textbf{SPCV2} and \textbf{AS-20K}, we use Mixup \cite{kong2020panns} and RRC for data augmentation. As for supervised downstream tasks, Mixup mixes both audio clips and labels. For \textbf{VOX1}, data augmentation is not applied. 
 %Note that, the Mixup methods used for pre-training and downstream tasks are different, the former only mixes the audio clips since labels are not involved in training, while the latter mixes both audio clips and labels. For
 
Classification accuracy (Acc) is taken as the performance metric for single-label tasks, including audio scene classification, spoken command recognition, speaker identification and music instrument classification, and mean average precision (mAP) for the task of multi-label sound event classification. For \textbf{US8K}, we conduct 10-fold cross-validation, and report the average accuracy of the 10 folds. 

\begin{figure}[t]
  \centering
  \includegraphics[width=\linewidth]{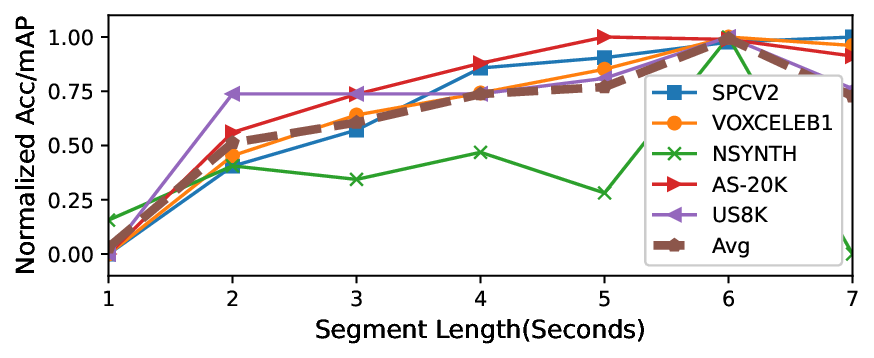}
  %\vspace{-6mm}
  \caption{Normalized Acc/mAP as a function of segment length, Acc/mAP of each task is normalized into the range of (0,1). "Avg" denotes averaging over normalized score of all tasks. }
  \label{fig:length}
\end{figure}

\subsection{Ablation study}
\label{sec:ablation}
 We separately evaluate the effectiveness of transformer encoder and the proposed view creation strategy. Ablation experiments are conducted using the \textbf{Small} model with the linear evaluation protocol, due to their low computational complexities. Table~\ref{tab:segments} shows the results. The result of BYOL-A is also given, which uses a CNN encoder and a single 1-second segment. Our models use a single or two segments, with a length of 1 second or 6 seconds. For a fair comparison, when the segment length is set to 1 second, we split audio clips into 1-second chunks for downstream tasks.
 
 \textbf{Transformer Encoder:} With the same view creation strategy, i.e. creating two views from a 1-second segment, our model (line 2 in Table \ref{tab:segments}) outperforms BYOL-A, especially for the two speech tasks ( \textbf{SPCV2} and \textbf{VOX1}). Speech involves more long-term semantic information, and transformer is more suitable than CNN for learning these long-term dependencies.
 
 \textbf{View Creation Strategy:} As shown in Table \ref{tab:segments}, when the segment length is set to 1 second, using one single segment is better than using two segments. This phenomenon is consistent with the claim made in BYOL-A \cite{niizumi_byol_2021} that the two segments may be too different to be identified as a positive pair. However, two views created from a single segment may share too much semantic content, thus leading our model to find an easy solution. 
When the segment length is increased to 6 seconds, the performance measures of \textbf{AS-20K}, \textbf{VOX1} and \textbf{US8K} are systematically increased, no matter whether using one or two segments. This is partially due to the capability of learning long-term dependencies of the transformer encoder.
In addition, for the 6-seconds case, using two segments exhibits superior performance over using one segment. The possible reasons are: the two segments can be rationally identified as a positive pair as they share a small portion of overlap, and meanwhile they are different enough to increases the task difficulty and thus leads the model to learn a more generalized representation.  
%Overall, the best average result is achieved by the proposed view creation strategy, namely using two 6 seconds segments.

%\vspace{-0.3cm}

%\vspace{-0.3cm}
Fig. \ref{fig:length} shows the normalized performance of each task as a function of segment length, where two segments are used. We can see that the performance measures increase along with the increasing of segment length until 6 seconds. This further verifies our new findings: i) when transformer encoder is used, increasing the segment length helps to learn more information; ii) when two segments are used, the segment length should be set
to make the segments share a proper amount of overlap, and have a proper difficulty for matching them as a positive pair.

\subsection{Linear Evaluation Results}

Table \ref{tab:le_result} shows the linear evaluation results. For fair comparison, we compare with other methods that also use Audioset for pre-training, including TRILL \cite{shor2020towards},  COLA \cite{saeed_contrastive_2020} and BYOL-A \cite{niizumi_byol_2021}. 
The performance scores are directly quoted from their original papers. It can be seen that our \textbf{Small} model outperforms other methods on all tasks, even though it only uses 1/10 data of Audioset-2M, while BYOL-A and COLA use the full Audioset-2M dataset. In particular, on the speaker identification task, our \textbf{Small} model obtains an accuracy of 61.9\%, compared to the 40.1\% accuracy of BYOL-A. By increasing the network size and the amount of training data, the performance can be further systematically increased by our \textbf{Base} model. The superiority of the proposed model comes from the use of transformer encoder and the proposed view creation strategy, which both are critical modules for the success of contrastive learning based pretraining technique. 

\def\a{true}
\def\b{false}
\setlength\tabcolsep{3pt}
\begin{table}[tb]
  \caption{Linear evaluation results.}
  \label{tab:le_result}
  \centering
  \footnotesize
     \scalebox{1}{
  \begin{threeparttable}
  \begin{tabular}{l|ccccc}
    \toprule
    \textbf{Method}   & \textbf{\makecell{AS-20K\\mAP}}    & \textbf{\makecell{SPCV2\\Acc (\%)}}    & \textbf{\makecell{VOX1\\Acc (\%)}}   
     & \textbf{\makecell{NSYNTH\\Acc (\%)}} & \textbf{\makecell{US8K\\Acc (\%)}}\\
    \midrule
    \textbf{TRILL \cite{shor2020towards} }  & -&-&17.9 &- &- \\
    \textbf{COLA \cite{saeed_contrastive_2020}} & - & $62.4$ &  $29.9$  & $63.4$   & -                                    \\
    \textbf{BYOL-A \cite{niizumi_byol_2021}}   & - & $92.2$ &  $40.1$  & $74.1$      & 79.1                                 \\
        \midrule
    \textbf{Small (ours)}  & 0.279 & 93.6 &  61.9  & 75.3 &   82.0                                    \\
    \textbf{Base (ours)} & \textbf{0.338} & \textbf{95.1} &  \textbf{72.0}  & \textbf{75.6}& \textbf{84.1}
                                      \\
    
    \bottomrule
  \end{tabular}
    %\begin{tablenotes}
  %\item{*} Speech containing clips in Audioset\cite{gemmeke_audio_2017}, roughly 1 million clips
  %\end{tablenotes}

  \end{threeparttable}
  }
\end{table}

\setlength\tabcolsep{3pt}
\begin{table}[tb]
  \caption{Finetuning results.}
  \label{tab:finetune_result}
  \centering
  \footnotesize
   \scalebox{1}{
  \begin{threeparttable}
  \begin{tabular}{l|c|cccc}
    \toprule
    \textbf{Method}    &  \textbf{\# Params} &\textbf{\makecell{AS-20K\\mAP}}    & \textbf{\makecell{SPCV2\\Acc (\%)}}    & \textbf{\makecell{VOX1\\Acc (\%)}}    
     \\
    \midrule
    \textbf{COLA \cite{saeed_contrastive_2020}}  &  & - & $95.5$ &  $37.7$                                      \\
    \textbf{Small-SSAST \cite{gong_ssast_2022}} & 23M   & 0.308 & $97.7$ &  $60.9$                                         \\
    \textbf{SSAST-PATCH \cite{gong_ssast_2022}}  &89M & 0.310 & $98.0$ &  $64.2$                                        \\
    \textbf{SSAST-FRAME \cite{gong_ssast_2022}}  & 89M  & 0.292 & \textbf{98.1} &  $80.8$                                        \\
    \textbf{Conformer \cite{srivastava_conformer-based_2022}} & 88M  & 0.276 & - & - \\
    \midrule
    \textbf{Small (ours)} & 22M   & 0.315 & $97.6$ &  $88.3$                                         \\
    \textbf{Base (ours)} & 86M  & \textbf{0.374} &  98.0 &  \textbf{94.3}  
                                      \\
    
    \bottomrule
  \end{tabular}
  %\begin{tablenotes}
  %\item{*} Librispeech Dataset\cite{panayotov2015librispeech}
  %\item{**} Self-hold dataset\cite{srivastava_conformer-based_2022}
  %\end{tablenotes}

  \end{threeparttable}
  }
\end{table}

\subsection{Fine-tuning Results}
To evaluate to what extent our models can further achieve, finetuning  experiments are conducted on the tasks of multi-label audio event classification (\textbf{AS-20K}), Spoken command recognition (\textbf{SPCV2}) and speaker identification (\textbf{VOX1}). We compare with those methods that also report the finetuning results, including  COLA \cite{saeed_contrastive_2020}, SSAST \cite{srivastava_conformer-based_2022} and Conformer \cite{srivastava_conformer-based_2022}. 
Table \ref{tab:finetune_result} shows the results. Compared to the best results achieved by other methods, our \textbf{Base} model performs better on  \textbf{AS-20K} (0.374 versus 0.310) and \textbf{VOX1} (94.3\% versus 80.8\%) by a large margin, and perform comparably on \textbf{SPCV2}. Remarkably, our \textbf{Small} model performs even better than SSAST and Conformer on \textbf{AS-20K} and \textbf{VOX1}, by using a much smaller network (22 M versus about 88 M). It is worth mentioning that both SSAST and Conformer use a transformer encoder as the proposed model, and use a wav2vec2-style pre-training scheme. Better results achieved by the proposed model may indicate that the teacher-student scheme is superior to the wav2vec2-style scheme for (segment-level) general audio pre-training.  %And it is reported  

\section{Conclusions}
\label{sec:conclusion}
In this work, we propose a general audio pre-training method with a  transformer-based teacher-student scheme, named ATST. A new view creation strategy is also proposed to fully leverage the capability of transformer. We evaluate the learned representation on diverse downstream tasks. Experiments show that the proposed view creation strategy is able to improve pre-training by properly increasing the difficulty of positive pair matching. Overall, the proposed model achieves the new state-of-the-art results on almost all of the tasks. We hope our work can facilitate the progress of general audio representation learning. 
%Future work includes extending downstream tasks, e.g. sound event detection, etc.

%\bibliographystyle{IEEEtran}

%\bibliography{mybib}

\end{document}